\documentclass{ws-ijgmmp}

\usepackage{color}

\def\cH{\mathcal{H}}

\def\PH{\mathbb{P}\mathcal{H}}
\def\e{\mathrm{e}}
\def\ii{\mathrm{i}}
\def\d{\mathrm{d}}

\newcommand{\ket}[1]{|{#1}\rangle}
\newcommand{\bra}[1]{\langle{#1}|}

\newcommand{\bracket}[2]{\langle#1|#2\rangle}

\numberwithin{equation}{section}

\begin{document}

\markboth{P. Facchi and S. Pascazio}
{Geometry of the quantum Zeno effect}

%
\catchline{}{}{}{}{}
%

\title{THE GEOMETRY OF THE QUANTUM ZENO EFFECT
}

\author{PAOLO FACCHI
}

\address{Dipartimento di Matematica and MECENAS, Universit\`a di Bari, 
I-70125 Bari,  Italy
\\
INFN, Sezione di Bari, I-70126 Bari , Italy\\
\email{paolo.facchi@ba.infn.it
} }

\author{SAVERIO PASCAZIO}

\address{Dipartimento di Fisica and MECENAS, Universit\`a di Bari,  I-70126 Bari, Italy\\
INFN, Sezione di Bari, I-70126 Bari, Italy\\
saverio.pascazio@ba.infn.it}

\maketitle

%
%

\begin{abstract}
The quantum Zeno effect is described in  geometric terms. 
The quantum Zeno time (inverse standard deviation of the Hamiltonian) and the generator of the quantum Zeno dynamics are both given a geometric interpretation.
\end{abstract}

\keywords{Quantum Zeno effect; Geometric quantum mechanics}

\section{Introduction and dedication}	
According to Wikipedia \cite{wiki}, a dedication ``refers to the inscription of books or other artifacts when these are specifically addressed or presented to a particular person. This practice, which once was used to gain the patronage and support of the person so addressed, is now only a mark of affection or regard. In law, the word is used of the setting apart by a private owner of a road to public use." 

We do neither intend to gain Beppe Marmo's support, nor show him our (long-standing) affection and regard. Rather, we shall adopt the afore-mentioned jurisdictional connotation of the word ``dedication". There is a road to understanding physics that makes use of geometry. Einstein was a precursor in the field. Beppe has been for a long time an outstanding advocate of such a geometric understanding. 
On many different occasions, we have discussed with him the basic notions of the quantum Zeno effect \cite{commonzeno}. A geometrical picture of such a fundamental quantum mechanical effect was proposed by Beppe at the conference for the 75th birthday of George Sudarshan \cite{7quests}, who fathered the quantum Zeno effect (QZE) in a seminal article written in collaboration with Baidyanath Misra \cite{MS77}. 

The objective of this note is to discuss this geometric road to the QZE, both in general terms and by making use of a concrete example.
In jurisdictional terminology, the ``dedication" is therefore \emph{by} Beppe to public use, and to the authors of this article in particular.
The effect of a dedication is that the freehold of the road is transferred to the public, who can avail themselves of it. In physics (unlike with roads)  this has a price, in that one has to learn how to speak the beautiful language of geometry. Beppe is aware of it. In fact we suspect that this was his original plan: to inculcate his friends with the love of the geometrical language and approach to quantum mechanics.

\section{Quantum Zeno effect and quantum Zeno dynamics}	
Consider a quantum system described in a Hilbert space  $\cH$,  whose dynamical evolution is governed by a Hamiltonian operator $H$. Let $\ket{\psi_0} \in \cH$ be its initial condition (pure state). The evolution is
\begin{equation}
\ket{\psi_t} = \e^{-\ii H t} \ket{\psi_0},
\end{equation}
with $\bracket{\psi_0}{\psi_0}=1, \bracket{\psi_t}{\psi_t}=1$.
The survival amplitude and probability read
\begin{equation}
A(t) = \bracket{\psi_0}{\psi_t}=\bra{\psi_0}\e^{-\ii H t}\ket{\psi_0}, \qquad
p(t)= |A(t)|^2 , 
\label{eq:survp}
\end{equation}
respectively.
At small times
\begin{equation}
p(t) \sim 1 - \frac{t^2}{\tau_{\mathrm{Z}}^2}.
\label{eq:ptauZ}
\end{equation}
The above quadratic behavior stems from first principles and very general features of the Schr\"odinger equation. The time constant 
\begin{equation}
\tau_{\mathrm{Z}}^{-2} = (\triangle H)^2 = 
\frac{\bra{\psi_0} H^2 \ket{\psi_0}}{\bracket{\psi_0}{\psi_0}} - \left(\frac{\bra{\psi_0} H \ket{\psi_0}}{\bracket{\psi_0}{\psi_0}}\right)^2
\label{eq:tauZ}
\end{equation}
is the \emph{Zeno time}. We explicitly wrote the norms in Eq.\ (\ref{eq:tauZ}) in order to make clear that the relevant objects belong to the complex projective space $\PH$.

Let $P$ be a projection onto $\cH_P= P\cH$ and $\ket{\psi_0} =P \ket{\psi_0}$ (state preparation). 
The evolution after $N$ projective measurements at time intervals $t/N$ reads
\begin{equation}
\ket{\psi_t^{(N)}} =  V_N(t) \ket{\psi_0}, \qquad
V_N(t)= V(t/N)^N = (P U(t/N) P)^N.
\label{eq:ZHN}
\end{equation}
Notice that $\ket{\psi_t^{(N)}}$ is not normalized.
We are interested in the $N \to \infty$ limit of the above evolution. The general case (unbounded $H$, infinite-dimensional $P$) is still open \cite{zenorev}. However, if $H$ is bounded and/or $P$ finite-dimensional, the limiting evolution is unitary and readily computed:
\begin{equation}
\ket{\psi_t^{(\textrm{Z})}} =  U_{\textrm{Z}}(t) \ket{\psi_0}, \qquad
U_{\textrm{Z}}(t) := \lim_{N \to \infty} V_N(t) = \e^{-\ii H_{\textrm{Z}}t} P, \qquad H_{\textrm{Z}} := PHP.
\label{eq:ZH}
\end{equation}
The (self-adjoint) generator $H_{\textrm{Z}}$ is called \emph{Zeno Hamiltonian} and the above equation defines the (unitary) \emph{quantum Zeno dynamics}.
Our main objectives will be to gain a geometrical intepretation of the Zeno time (\ref{eq:tauZ}) and of the Zeno Hamiltonian (\ref{eq:ZH}).
An attempt to interpret the Zeno time in geometrical terms was done in \cite{PL98}.

\section{Geometry of quantum mechanics}

States are not vectors $| \psi \rangle \in \mathcal{H}$ but rather rays, elements of the Hilbert manifold $\PH$, which can conveniently parametrized as rank-one projection operators.
This projection map enables one to identify on $\PH$ a metric tensor, usually called the Fubini-Study metric, and a symplectic structure \cite{marmo1,marmo3}. Both of them define on $\PH$ what is called a K\"ahlerian structure. 
To work with this tensor on $\mathcal{H}$ instead of $\PH$ is quite convenient for computational purposes. 

We shall consider finite dimensional systems. Let $\cH \simeq \mathbb{C}^n$
and introduce the orthornormal basis $\{\ket{e_1}, \dots, \ket{e_n}\}$ and coordinates 
$z_k = (q_k + \ii p_k)$. This entails
\begin{equation}
\cH \to \mathbb{C}^n \to \mathbb{R}^{2n},
\end{equation}
with
\begin{equation}
\ket{\psi} \mapsto 
\left(\begin{array}{c}  \bracket{e_1}{\psi} \\\bracket{e_2}{\psi} \\ \vdots \\ \bracket{e_n}{\psi}\end{array}\right)
=: \left(\begin{array}{c}  z_1 \\ z_2 \\ \vdots \\ z_n \end{array}\right)
=: \left(\begin{array}{c}  q_1+\ii p_1 \\ q_2+\ii p_2 \\ \vdots \\ q_n + \ii p_n \end{array}\right)
\mapsto \left(\begin{array}{c}  q_1 \\ \vdots \\ q_n \\  p_1 \\ \vdots \\ p_n  \end{array}\right) .
\end{equation}
The complex inner product of $\cH$ induces the Hermitian structure
\begin{equation}
\mathfrak{h} = \sum_k   \d \bar{z}_k \otimes \d z_k = \mathfrak{g} + \frac{\ii}{2} \omega,
\end{equation}
where 
\begin{eqnarray}
\mathfrak{g}&=& \frac{1}{2} \sum_k (\d \bar{z}_k \otimes \d z_k +  \d z_k  \otimes \d \bar{z}_k )
=  \sum_k  \d \bar{z}_k \otimes_s \d z_k , \\
\omega &=& - \ii  \sum_k (\d \bar{z}_k \otimes \d z_k - \d z_k  \otimes \d \bar{z}_k )
= - \ii  \sum_k \d \bar{z}_k \wedge \d z_k  
\end{eqnarray}
are a Riemannian metric and a symplectic form, respectively. In terms of the real coordinates, they read
\begin{eqnarray}
\mathfrak{g} = \sum_k (\d q_k \otimes \d q_k + \d p_k \otimes \d p_k ),  \qquad 
\omega = \sum_k \d q_k \wedge \d p_k ,
\end{eqnarray}
respectively.
Let us introduce the dual  coordinate vector fields
\begin{equation}
\frac{\partial}{\partial z_k} = \frac{1}{2} \left(\frac{\partial}{\partial q_k} - \ii \frac{\partial}{\partial p_k}\right), \quad
\d z_j \left(\frac{\partial}{\partial z_k}\right) = \frac{\partial}{\partial z_k} z_j = \delta^k_j, \quad
\d \bar{z}_j \left(\frac{\partial}{\partial z_k}\right) = 0
\end{equation}
and analogously for $\frac{\partial}{\partial \bar{z}_k}$. 
In terms of them we can write the contravariant form of the Hermitian structure:
\begin{equation}
K = \sum_k \frac{\partial}{\partial \bar{z}_k} \otimes \frac{\partial}{\partial z_k} = G + \frac{\ii}{2} \Omega ,
\end{equation}
where
\begin{eqnarray}
G &=& \sum_k \frac{\partial}{\partial \bar{z}_k} \otimes_s \frac{\partial}{\partial z_k}  =\frac{1}{4}
\sum_k \left(\frac{\partial}{\partial q_k} \otimes \frac{\partial}{\partial q_k} + \frac{\partial}{\partial p_k} \otimes \frac{\partial}{\partial p_k} \right) , \\
\Omega &=& -\ii \sum_k \frac{\partial}{\partial \bar{z}_k} \wedge \frac{\partial}{\partial z_k}
= - \frac{1}{2} \sum_k  \frac{\partial}{\partial q_k} \wedge \frac{\partial}{\partial p_k} .
\end{eqnarray}
By using $\Omega$ we can construct the Hamiltonian vector field $X_f$ generated by any (smooth) Hamiltonian function $f$, in the standard way:
\begin{equation}
X_f = \Omega (\d f , \cdot ) =
- \ii \sum_k \left(\frac{\partial f }{\partial \bar{z}_k}  \frac{\partial}{\partial z_k} - \frac{\partial f}{\partial z_k}  \frac{\partial}{\partial \bar{z}_k} \right)=
\frac{1}{2} \sum_k \left(\frac{\partial f }{\partial p_k}  \frac{\partial}{\partial q_k} - \frac{\partial f}{\partial q_k}  \frac{\partial}{\partial p_k} \right). 
\label{eq:XH}
\end{equation}
It is easy to verify that it satisfies the equation $\omega(X_f,\cdot) = \d f$. The tensor $\Omega$ defines also a Lie algebra structure on functions, with  the Poisson brackets
\begin{eqnarray}
\{ f, g\} &=&  \Omega(\d f, \d g) 
\nonumber\\
&=&-\ii \sum_k \left(\frac{\partial f }{\partial \bar{z}_k}  \frac{\partial g}{\partial z_k} - \frac{\partial f}{\partial z_k}  \frac{\partial g}{\partial \bar{z}_k} \right)
=\frac{1}{2}\sum_k \left(\frac{\partial f }{\partial p_k}  \frac{\partial g}{\partial q_k} - \frac{\partial f}{\partial q_k}  \frac{\partial g}{\partial p_k} \right).
\end{eqnarray}
On the other hand, $G$ defines a Jordan algebra on all quadratic functions, with
the Jordan brackets
\begin{eqnarray}
\{ f, g\}_+  &=&    G(\d f, \d g) 
\nonumber\\
&=& \frac{1}{2} \sum_k \left(\frac{\partial f }{\partial \bar{z}_k}  \frac{\partial g}{\partial z_k} + \frac{\partial f}{\partial z_k}  \frac{\partial g}{\partial \bar{z}_k} \right) =
\frac{1}{4}\sum_k \left(\frac{\partial f}{\partial q_k}  \frac{\partial g}{\partial q_k} + \frac{\partial f}{\partial p_k}  \frac{\partial g}{\partial p_k} \right) .
\end{eqnarray}
The real quadratic functions are the expectation values of observables
\begin{equation}
f_A(\psi)=\bra{\psi}A \ket{\psi} = \sum_{k,l} \bar{z}_k A_{kl} z_l.
\label{eq:expval}
\end{equation}
$(\Omega, G)$ defines a Lie-Jordan algebra on these expectation values,
\begin{equation}
\label{eq:Lie-Jordan}
\{f_A, f_B\} = f_{\ii (AB-BA)}, \qquad
\{f_A, f_B\}_+ = f_{\frac{1}{2}(AB+BA)},
\end{equation}
isomorphic to the Lie-Jordan algebra of the observables.

The Hamilton equations, whose Hamiltonian vector fields~(\ref{eq:XH}) are generated by quadratic functions $f_H = \bra{\psi} H \ket{\psi}$,
\begin{equation}
\dot{z}_k = X_{f_H} (z_k) = -\ii \frac{\partial f}{\partial \bar{z}_k} = -\ii \sum H_{k l } z_l,
\end{equation}
are nothing but the Schr\"odinger equation with Hamiltonian operator $H$,
$\dot{\psi} = H \psi$.
The evolution of any expectation value (\ref{eq:expval})
 is given in terms of the Poisson brackets
\begin{equation}
\dot{f}_A = \{f_H, f_A\} = f_{\ii (HA-AH)},
\end{equation}
which is the Heisenberg equation for the operator $A$,
$\bra{\psi} \dot{A} \ket{\psi} = \bra{\psi} \ii (HA-AH) \ket{\psi}$.

Projectability onto $\PH$ requires a conformal factor
\begin{equation}
\tilde{G}(\psi) = \bracket{\psi}{\psi} G(\psi), 
\qquad
\tilde{\Omega}(\psi) = \bracket{\psi}{\psi} \Omega(\psi).
\end{equation} 
Observe that $\tilde{\Omega}$ does not satisfy the Jacobi identity any more.
However, it does on expectation values. Thus, $\tilde{\Omega}$ projects onto $\PH$, a Poisson space. 
Moreover, one should consider homogeneous expectation values
\begin{equation}
\tilde{f}_A = \frac{\bra{\psi} A \ket{\psi} }{\bracket{\psi}{\psi}}.
\end{equation}

We are now ready to make the first key observation: the Zeno time (\ref{eq:tauZ}) is the inverse length of the Hamiltonian vector field on the complex projective space:
\begin{equation}
\tau_{\mathrm{Z}}^{-2} = \tilde{G} (\d \tilde{f}_H, \d \tilde{f}_H).
\end{equation}
Indeed, we have
\begin{equation}
\d \tilde{f}_H = \frac{1}{\|\psi\|^2} \left(\d f_H - \tilde{f}_H \d \|\psi\|^2\right),
\end{equation}
and from~(\ref{eq:Lie-Jordan}) we get
\begin{equation}
G(\d f_H, \d f_H) = \{f_H, f_H\}_+ = f_{H^2} = \bra{\psi} H^2 \ket{\psi}.
\end{equation}
Therefore,
\begin{eqnarray}
\tilde{G}(\d \tilde{f}_H, \d \tilde{f}_H) &=& \frac{1}{\|\psi\|^2} G\left( \d f_H - e_H \d \|\psi\|^2, \d f_H - e_H \d \|\psi\|^2\right)
\nonumber\\
&=& \tilde{f}_{H^2} - (\tilde{f}_H)^2 = (\triangle H)^2.
\end{eqnarray}

\section{An example: the qubit}
Let $\cH \simeq \mathbb{C}^2 \simeq \mathbb{R}^4$ with Hamiltonian
\begin{equation}
H = h_0 \mathbb{I} + \vec{h}\cdot\vec{\sigma} ,
\end{equation}
where $\vec{\sigma}=(\sigma_x,\sigma_y,\sigma_z)$ is the vector of Pauli matrices, $h_0 \in \mathbb{R}$, and $\vec{h}=(h_x,h_y,h_z)\in \mathbb{R}^3$. We introduce coordinates $z_1,z_2$ in the eigenbasis of $\sigma_z$, i.e.\ $\sigma_z \ket{e_j} = (-1)^j \ket{e_j}$, $j=1,2$.
The expectation value~(\ref{eq:expval}) of the Hamiltonian $H$ reads
\begin{eqnarray}
f_H = \bra{\psi}H \ket{\psi} 
= h_0 u +h_x x + h_y y + h_z z =  h_0 u + \vec{h}\cdot \vec{x},
\label{eq:fH}
\end{eqnarray}
where the quadratic functions $u, x, y, z$ are the expectation values of the Pauli matrices
\begin{eqnarray}
u &=& f_{\mathbb{I}} = |z_1|^2 + |z_2|^2 = q_1^2 + p_1^2 + q_2^2 + p_2^2,
\nonumber\\
x &=& f_{\sigma_x} = \bar{z}_1 z_2 + \bar{z}_2 z_1 = 2 (q_1 q_2 + p_1 p_2),
\nonumber\\
y &=& f_{\sigma_y} = -\ii(\bar{z}_1 z_2 -\bar{z}_2 z_1) = 2 (q_1 p_2 - p_1 q_2),
\nonumber\\
z &=& f_{\sigma_z} = |z_1|^2 - |z_2|^2 = q_1^2 + p_1^2 - q_2^2 - p_2^2.
\label{eq:fH4}
\end{eqnarray}
Notice that $u$ and $\vec{x}$ are not independent, since 
\begin{equation}
\label{eq:u-x2}
u^2 = \left(|z_1|^2 + |z_2|^2 \right)^2 = x^2 + y^2 + z^2 = \vec{x}^2.
\end{equation}
By making use of~(\ref{eq:u-x2}) we easily get that the Zeno time~(\ref{eq:tauZ}) is
\begin{equation}
\tau_{\mathrm{Z}}^{-2} = \tilde{G} (\d \tilde{f}_H, \d \tilde{f}_H) = \tilde{f}_{H^2} - (\tilde{f}_H)^2 = \vec{h}^2 - \frac{(\vec{h}\cdot \vec{x})^2}{u^2} = \frac{(\vec{h}\wedge \vec{x})^2}{\vec{x}^2}, 
\end{equation}
where $\wedge$ is the vector product in $\mathbb{R}^3$.

\begin{figure}[ph]
\centerline{\includegraphics[width=1.7in]{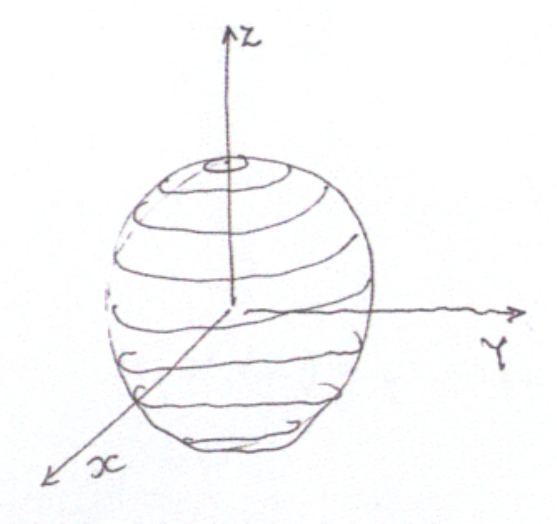}}
\vspace*{8pt}
\caption{Zeno flow of a qubit subject to measurement $P = \frac{1}{2}(\mathbb{I} + \sigma_3)$, according to Beppe Marmo, author of the drawing.}
\label{fig1}
\end{figure}
Let us now consider the Zeno effect with projection 
\begin{equation}
P = \ket{e_1}\bra{e_1}=\frac{1}{2}(\mathbb{I} + \sigma_3).
\label{eq:Pzeno}
\end{equation}
The Zeno dynamics  (\ref{eq:ZH}) becomes
\begin{equation}
U_{\textrm{Z}}(t) = \e^{-\ii H_{\textrm{Z}}t} P, \qquad
H_{\textrm{Z} }= P H P =   \frac{h_0 + h_z}{2}(\mathbb{I} + \sigma_3).
\label{eq:Pzenodyn}
\end{equation}
The Zeno Hamiltonian function reads
\begin{equation}
f_{H_{\mathrm{Z}}} =  \frac{h_0 + h_z}{2} (u + z) = (h_0 + h_z) |z_1|^2 = (h_0 + h_z) (q_1^2 + p_1^2),
\end{equation}
which is to be compared to the ``free" Hamiltonian function 
(\ref{eq:fH})-(\ref{eq:fH4}). 
The infinitesimal generator of the Zeno dynamics is
\begin{equation}
X_{\mathrm{Z}} 
= -\ii (h_0 + h_z) \left(z_1  \frac{\partial}{\partial z_1} -  \bar{z}_1  \frac{\partial}{\partial \bar{z}_1}\right)=
2 (h_0 + h_z)  \left( p_1 \frac{\partial}{\partial q_1} - q_1 \frac{\partial}{\partial p_1}\right)
\end{equation}
and the dynamics reads
\begin{equation}
\dot{u} =0, \quad \dot{x}= -2 (h_0 + h_z) y, \quad \dot{y}=  2 (h_0 + h_z) x, \quad \dot{z} =0.
\end{equation}
Observe that, due to (\ref{eq:u-x2}),
the first equation, $\dot{u} =0$, entails that the flow is on the sphere $|\vec x| =$ const.
This flow is depicted in Fig.~\ref{fig1}.
 
Notice however that the Zeno dynamics (\ref{eq:Pzenodyn}) implies, through the projection (\ref{eq:Pzeno}), also a constraint on the initial condition [state preparation, see the two lines that precede Eq.\ (\ref{eq:ZHN})]
\begin{equation}
z_2=0 \quad \Leftrightarrow \quad  x=y=0, \; z>0.
\label{eq:z2}
\end{equation}
Among all possible orbits, the constraint (\ref{eq:z2}) ``picks" the degenerate orbit on the North Pole and ``freezes" the dynamics in the initial state. This is the customary, simplest version of the quantum Zeno effect, obtained through a one-dimensional projection like (\ref{eq:Pzeno}).
On larger spaces, with multi-dimensional projections, one can obtain non-trivial interesting dynamics~\cite{zenorev}.

\section{Conclusions}
We have discussed the quantum Zeno effect and the quantum Zeno dynamics in geometric terms, by proposing a geometric intepretation of the Zeno time (\ref{eq:tauZ}) and of the Zeno Hamiltonian (\ref{eq:ZH}). The general scheme has been corroborated by the simplest non-trivial quantum-mechanical example, that of a spin-$1/2$ particle.
Beppe, who has always considered examples ``time consuming" (generalization being the main road towards ``understanding"), will hopefully forgive us: the example is his own and it helped us understand. Further generalizations of the scheme described in this note have \emph{already} been considered by Beppe and will be published elsewhere.

\section*{Acknowledgments}
We thank Beppe Marmo for many conversations on the physical and geometrical aspects of the quantum Zeno effect.

\end{document}